\documentclass[aps,floatfix,nofootinbib,preprint]{revtex4}
\usepackage{amsmath}
\usepackage{epsfig}
\usepackage{color}
\usepackage{endnotes}
\let\footnote=\endnote

\newcommand{\be}{\begin{equation}}
\newcommand{\ee}{\end{equation}}

\newcommand{\ber}{\begin{eqnarray}}
\newcommand{\eer}{\end{eqnarray}}

\begin{document}

\title{Coupling the Lorentz Integral Transform (LIT) and the Coupled Cluster (CC)
Methods: A way towards continuum spectra of ``not-so-few-body''
systems.}

\author{Giuseppina Orlandini$^{1,2}$, Sonia Bacca$^{3,4}$,  Nir Barnea$^{5}$, Gaute Hagen$^{6,7}$,  
        Mirko Miorelli$^{1,3}$, Thomas Papenbrock$^{7,6}$}
\affiliation{
$^{1}$ Dipartimento di Fisica, Universit\`a di Trento, I-38123 Trento, Italy\\
$^{2}$ Istituto Nazionale di Fisica Nucleare, Gruppo Collegato di Trento, I-38123 Trento, Italy\\
$^{3}$ TRIUMF, 4004 Wesbrook Mall, Vancouver, BC, V6T 2A3, Canada\\
$^{4}$ Department of Physics and Astronomy, University of Manitoba, Winnipeg, MB, R3T 2N2, Canada \\
$^{5}$ Racah Institute of Physics, The Hebrew University, 91904, Jerusalem, Israel\\
$^{6}$ Physics Division, Oak Ridge National Laboratory, Oak Ridge, TN 37831, USA \\
$^{7}$ Department of Physics and Astronomy, University of Tennessee, Knoxville, TN 37996, USA 
}

\begin{abstract}
Here we summarize how the LIT and CC methods can be coupled, in order to allow for {\it ab initio} calculations of
reactions in medium mass nuclei. Results on $^{16}$O are reviewed and preliminary calculations 
on $^{40}$Ca are presented.
\end{abstract}

\maketitle

\section{Introduction}
\label{intro}
A striking characteristic of the nuclear photo-absorption cross section is a very pronounced peak around
10--20~MeV called the giant dipole resonance (GDR). Historical semi-classical interpretations are based on
a dipole collective motion of protons against neutrons~\cite{Ref1}. Later, both collective 
and in microscopic many-body approaches have tried to account for its centroid and width~\cite{Ref2,Ref3}. 
For {\it ab initio} approaches the difficulty is the position of the resonance in the continuum part of the spectrum. 

The LIT method~\cite{Ref4} reduces the continuum problem into a bound state problem. Its application however,
has been restricted to $A<8$ nuclei~\cite{Ref5}. Therefore it is expedient to try to couple it with the 
CC method~\cite{Ref6,Ref7}, which is very powerful in dealing with bound states of many-body systems.  
\section{Formalism}
\label{sec:1}
The main point of the LIT method is that the function $S(\omega)$ entering the photonuclear cross 
section ($\sigma_\gamma(\omega)=4\,\pi^2\,\alpha \,\omega S(\omega)$)
can be accessed via inversion of its integral transform with a Lorentzian kernel
\begin{equation}\label{eq:(1)}
L(\omega_0,\Gamma)\ =\  \frac{\Gamma}{\pi}\int d\omega\ \frac{S(\omega)} {(\omega -\omega_0)^2 + \Gamma^2}\,,
\end{equation}
In dipole approximation $S(\omega)$ is given by
\begin{equation}
 S(\omega)= \sum_n|\langle \Psi_n| {\bf D} |\Psi_0\rangle|^2\delta(\omega-E_n+E_0)\,,
\end{equation}
where $\omega$ is the photon energy and $E_0,E_n$ and $|\Psi_0\rangle,|\Psi_n\rangle$ are ground and
excited state eigenvalues and eigenstates, respectively.
The completeness property of the hamiltonian eigenstates allows to write
\begin{equation}\label{eq:(2)}
L(\omega_0,\Gamma)=\langle\Psi_0| {\bf D} \frac{1}{(H-E_0-\omega_0-i \Gamma)}
\frac{1}{(H-E_0-\omega_0+ i \Gamma)} {\bf D}|\Psi_0\rangle\equiv
\langle{\tilde \Psi}|{\tilde \Psi}\rangle\,.  
\end{equation}
Since the integral in (1) is finite $ |{\tilde \Psi}\rangle$ is bound, therefore bound state techniques can 
be used to calculate its norm $L(\omega_0,\Gamma)$, for fixed values of $\Gamma$ and many values of $\omega_0$. 
For $A<8$ hyperspherical harmonics (HH) expansions have been used~\cite{Ref8}. 
Accurate enough results for $L$ have allowed its inversion~\cite{Ref9,Ref10} to give the required response function.
Here we reformulate the LIT  within the CC approach. To this aim we rewrite Eq. (3) introducing 
a similarity transformation $ e^T $: 
\begin{eqnarray}\label{eq:(3)}
L(\omega_0,\Gamma)&=&\langle \Psi_0|e^Te^{-T}   {\bf D}e^T e^{-T} \frac{1}{(H-E_0-\omega_0-i\Gamma)}e^Te^{-T} 
\frac{1}{(H-E_0-\omega_0+ i\Gamma)}  e^T e^{-T} {\bf D}e^T e^{-T} |\Psi_0\rangle \nonumber\\ 
&\equiv&\langle0_L|{\bf \bar D} \frac{1}{(\bar H-E_0-\omega_0-i \Gamma)} 
\frac{1}{(\bar H-E_0-\omega_0+ i \Gamma)}  
{\bf \bar D} |0_R\rangle \equiv \langle\bar \Psi_L|\bar\Psi_R\rangle
\end{eqnarray} 
Notice that $\langle 0_L|$ is different from $|0_R\rangle$.
In the CC approach the operator $T$ is such that $|0_R\rangle\equiv e^{-T} |\Psi_0\rangle$ is a single Slater 
determinant, and in the single-double (CCSD) approximation it is chosen as a linear combination  
of one-particle-one-hole (1p-1h) and two-particle-two-hole (2p-2h) operators, only. The amplitudes of $T$ are obtained by solving 
the CC equations~\cite{Ref11}.

To calculate the LIT within the CC method one has the additional problem to find $|\bar\Psi_R\rangle$ 
and $\langle\bar \Psi_L|$, which are solutions of the following equations:
\begin{eqnarray}\label{eq:(4)}
\langle\bar \Psi_L| (\bar H-E_0-\omega_0-i \Gamma)&=&\langle0_L|{\bf \bar D}\,; \nonumber \\
(\bar H-E_0-\omega_0+i \Gamma) |\bar\Psi_R\rangle&=&  {\bf \bar D} |0_R\rangle \,.
\end{eqnarray}
To this aim one may use the equation of  motion (EoM) method, namely one can write $|\bar\Psi_R\rangle = {\cal R} |0_R\rangle$
and $\langle\bar \Psi_L|= \langle 0_L| {\cal L}\,,$ 
%\begin{equation} \label{eq:(5)}
%|\bar\Psi_R\rangle = {\cal R} |0_R\rangle \,\,;\,\,\langle\bar \Psi_L|= \langle 0_L| {\cal L}\,,  
%\end{equation}
where ${\cal R}$ and ${\cal L}$ are linear combinations of 1p-1h and 2p2h
excitation operators, similar to $T$. Their amplitudes are obtained by solving the following  equations 
\begin{eqnarray} \label{eq:(6)}
   [\bar{H},{\cal R}]\,|0_R\rangle &=& (\omega_0-i\Gamma) {\cal R} |0_R\rangle \nonumber+\bar{D}|0_R \rangle\,;\\
 \langle 0_L\,[\bar{H},{\cal L}] &=& \langle 0_L |{\cal L}(\omega_0+i\Gamma)+\langle 0_L\bar{D}\,.
 %+\langle 0_L\bar{D}\.
\end{eqnarray}
They differ from the CC EoM only by the presence of the source terms 
$\bar{D}|0_R \rangle$ and $\langle 0_L|\bar{D}$.

\section{Results}
\begin{figure*}[htb]
\centering
\includegraphics[width=0.8\textwidth]{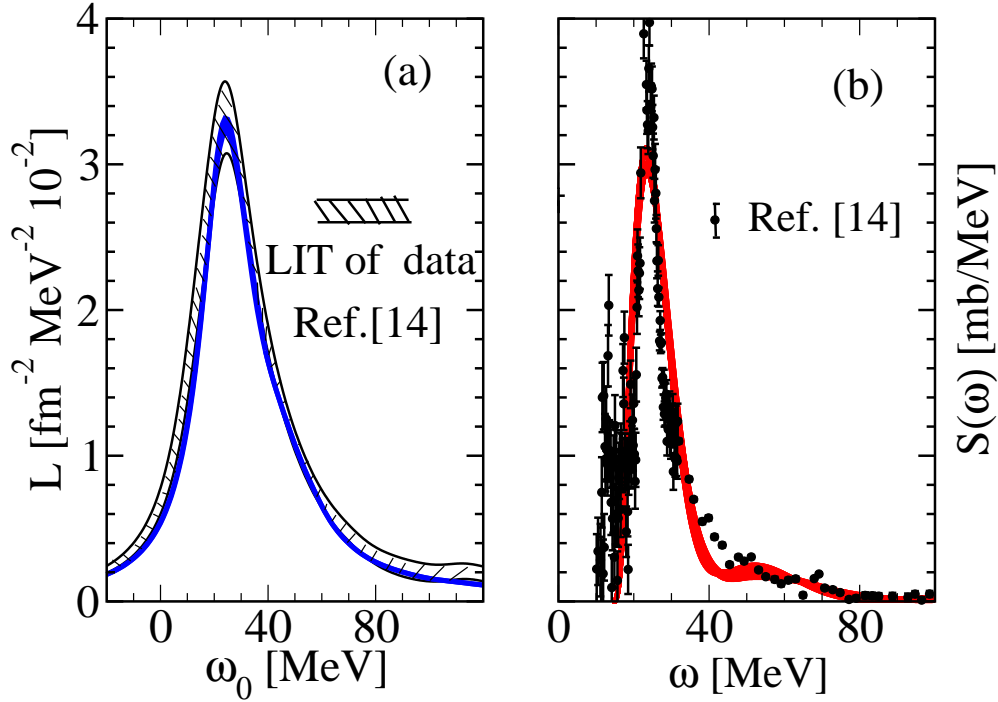}
\caption{(Color online) (a): Comparison of the LIT at $\Gamma=10$~MeV for $N_{max}=18$  
and the Lorentz integral transform of Ahrens {\it et al.}~data~\cite{Ref14}. (b): Comparison
of $S(\omega)$ of $^{16}$O dipole response against experimental data.}
\label{Fig1}
\end{figure*}

\begin{figure*}[htb]
\centering
\includegraphics[width=0.8\textwidth]{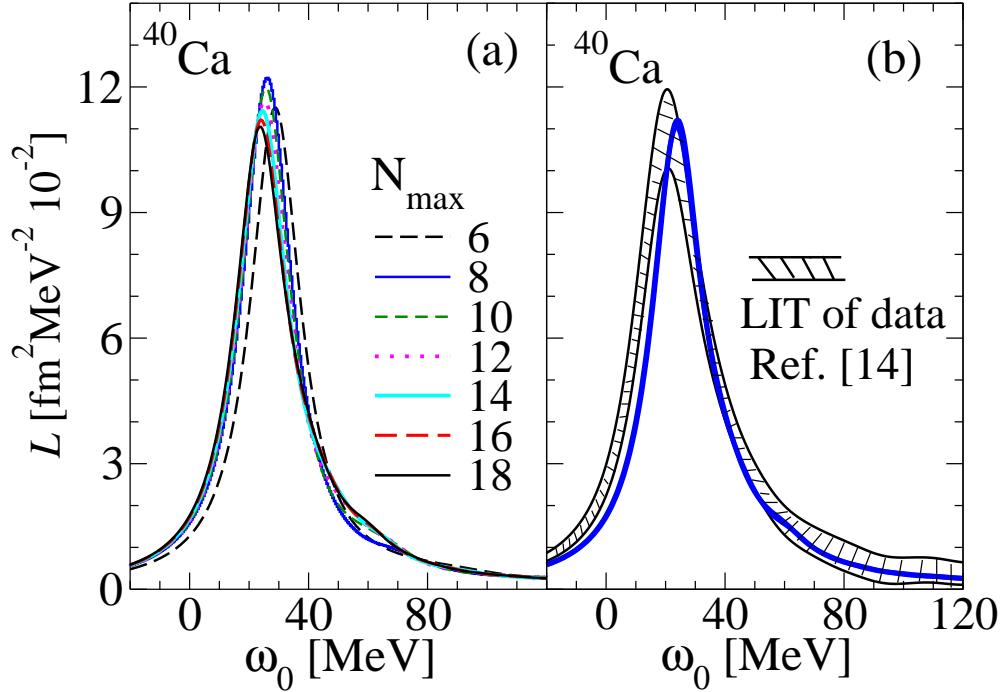}
\caption{(Color online) (a): Convergence of  $L(\omega_0,\Gamma)$ at
  $\Gamma=10$~MeV as a function of $N_{max}$. (b) Comparison of the LIT at $\Gamma=10$~MeV for $N_{max}=16,18$  
  and the LIT of Ahrens {\it et al.}~data~\cite{Ref14}.}
\label{Fig2}
\end{figure*}

The results presented in the following have been obtained using a realistic chiral effective field theory potential 
(N3LO~\cite{Ref12}). The method  has been first validated on the $S(\omega)$ of $^4$He. 
This has been obtained from the inversion of the LIT calculated by expanding $|\tilde\Psi\rangle$ 
%(see Eq.~(\ref{eq:(2)})) 
in HH, up to full convergence. Therefore the HH calculation can be considered virtually 
``exact'' and the excellent agreement with the LIT-CC result can be seen in Ref.~\cite{Ref13}.
From the same reference we report in Fig.~\ref{Fig1} the results for $^{16}$O. 
In Fig.~\ref{Fig1}a the theoretical LIT is compared to the LIT of the data~\cite{Ref14}. The value of $\Gamma$ 
has been chosen to be 10 MeV, since it is the smallest value for which we are able to obtain a convergent result
(the smaller the value of $\Gamma$, the slower the rate of convergence). Nevertheless one can notice that 
the transform of the data maintains the resonant structure and preserves the peak position. This is due to 
the fact that the Lorentzian kernel is a representation of the delta-function (for $\Gamma=0$ the transform 
coincides with the response). While the comparison shows that the experimental centroid of the GDR 
is well reproduced by a calculation that neglects three-body forces, the inversion of the transform
leads to a less pronounced peak with respect to experiment (see Fig.~\ref{Fig2}b).
\bigskip\bigskip\bigskip

Aiming at addressing {\it ab initio} the interesting case of $^{48}$Ca, for which recent photoabsorption
measurements have been performed~\cite{Ref15}, we have calculated  the LIT of its 
N=Z  partner $^ {40}$Ca. In Fig.~\ref{Fig2}a we present preliminary results about the rate of convergence 
of the transform. As for $^{16}$O, in Fig.~\ref{Fig2}b we show the comparison with the LIT of the data~\cite{Ref14}. 
Three remarks are in order here:
\begin{itemize}
\item as can be seen in Fig.~\ref{Fig2}a full convergence is not yet reached, therefore an inversion is not worth; 
\item  different from the case of $^{16}$O a purely experimental comparison between the peak positions of 
the data in~\cite{Ref14} and of their LIT in Fig ~\ref{Fig2}b shows a slight difference. 
This is due to the fact that $S(\omega)$ has a tail at higher energies. This tail contributes to the LIT, 
which differs enough from a delta-function to shift its peak to the right. Therefore the LIT peak position 
cannot be considered a prediction of the centroid of the GDR, but one needs an inversion;
\item the use of larger model spaces seem to move the theoretical result towards the data.
However, the agreement in the transforms would not clearly imply an agreement 
in $S(\omega)$ (as the results on $^{16}$O have also shown), being only a minimal condition. 
On the other hand, a disagreement in the integral transforms already might give a important 
information, pointing to possible shortcomings in the potential (or in the neglect of higher clusters).
\end{itemize}                                                                                                
\section{Conclusions}
Here we have summarized the CC-LIT method, which allows to calculate response functions in 
the continuum  of not-so-few-body systems. We have presented {\it ab initio} results obtained with this method 
and using  only the chiral N3LO two-body potential. The application to the GDR of $^{16}$O has shown 
the ability of this potential to reproduce the centroid of the resonance, while a somewhat less 
pronounced structure has been found inverting the transform. Preliminary results on the application of 
the method to the GDR of $^{40}$Ca show that larger model spaces are still needed 
to reach a convergent LIT. There are indications that larger and larger model spaces might lead to an 
agreement between theoretical and experimental LIT's. Convergent results are needed to attempt an 
inversion of the transform.
\section{Acknowledgements}
This work was supported by the MIUR grant
PRIN-2009TWL3MX, the Natural Sciences and Engineering Research Council, the National Research
Council of Canada, the Israel Science Foundation (Grant number 954/09),  
the US-Israel Binational Science Foundation (Grant No 2012212), the Office of Nuclear
Physics, U.S. Department of Energy (Oak Ridge National Laboratory) and DE-SC0008499 
(NUCLEI SciDAC collaboration). Computer time was provided by the
Innovative and Novel Computational Impact on Theory and Experiment
(INCITE) program. This research used resources of the Oak Ridge
Leadership Computing Facility located in the Oak Ridge National
Laboratory, which is supported by the Office of Science of the
Department of Energy under Contract No. DE-AC05-00OR22725, and used
computational resources of the National Center for Computational
Sciences, the National Institute for Computational Sciences.

\end{document}